\documentclass[prb,amsmath,showpacs,twocolumn,floatfix]{revtex4-1}

\usepackage{amssymb}
\usepackage{graphicx}
\usepackage{subfigure}
\usepackage{color}
\usepackage{multirow}
\usepackage{verbatim}

\begin{document}

\title{Domain walls in a perovskite oxide with two primary structural
  order parameters: first-principles study of BiFeO$_3$}

\author{Oswaldo Di\'eguez,$^{1,2}$ Pablo Aguado-Puente,$^3$ Javier
  Junquera,$^3$ and Jorge \'I\~niguez$^1$}

\affiliation{$^1$Institut de Ci\`encia de Materials de Barcelona
  (ICMAB-CSIC), Campus UAB, 08193 Bellaterra, Spain\\ $^2$Department
  of Physics and Astronomy, Rutgers University, Piscataway, New Jersey
  08854, USA\\
  $^3$Departamento de Ciencias de la Tierra y
              F\'{\i}sica de la Materia Condensada, Universidad de Cantabria,
              Cantabria Campus Internacional,
              Avenida de los Castros s/n, 39005 Santander, Spain}

\begin{abstract}
We present a first-principles study of ferroelectric domain walls
(FE-DWs) in multiferroic BiFeO$_3$ (BFO), a material in which the FE
order parameter coexists with anti-ferrodistortive (AFD) modes
involving rotations of the O$_6$ octahedra. We find that the
energetics of the DWs are dominated by the capability of the domains
to match their O$_6$ octahedra rotation patterns at the plane of the
wall, so that the distortion of the oxygen groups is minimized. Our
results thus indicate that, in essence, it is the discontinuity in the
AFD order parameter, and not the change in the electric polarization,
what decides which crystallographic planes are most likely to {\em
  host} BFO's FE-DWs. Such a result clearly suggests that the O$_6$
rotational patterns play a primary role in the FE phase of this
compound, in contrast with the usual (implicit) assumption that they
are subordinated to the FE order parameter. Our calculations show that, for
the most favorable cases in BFO, the DW energy amounts to a several
tens of mJ/m$^2$, which is higher than what was computed for other
ferroelectric perovskites with no O$_6$ rotations. Interestingly, we
find that the structure of BFO at the most stable DWs resembles the
atomic arrangements that are characteristic of low-lying (meta)stable
phases of the material. Further, we argue that our results for the DWs
of bulk BFO are related with the nanoscale-twinned structures that
Prosandeev {\em et al}. [Adv.\ Funct.\ Mats.\ (2012), doi:
  10.1002/adfm.201201467] have recently predicted to occur in this
compound, and suggest that BFO can be viewed as a polytypic
material. Our work thus contributes to shape a coherent picture of the
structural variants that BFO can present and the way in which they are
related.
\end{abstract}



\maketitle


\section{Introduction}

Ferroelectric (FE) crystals are insulators displaying a macroscopic
polarization that can be switched by the action of an electric field.
The ground state of a bulk ferroelectric, where a unit cell is periodically
repeated in space, is a perfectly ordered
structure where the atomic patterns that give rise
to the polarization are the same in every part of the
crystal. However, real ferroelectric crystals need to minimize the
electrostatic energy arising from their finite size (i.e., the
electric field outside a ferroelectric sample must rapidly decay to
zero), and as a result the formation of regions where the polarization
points in different directions will occur. Like in the case of other
ferroic materials such as ferromagnets or ferroelastics, these regions
are called domains, and the area in between domains is called domain
wall (DW).

The presence of DWs in a material can cause important changes to its
macroscopic behavior.\cite{Salje2010CPC} Thus, for example, recent
findings about the conducting\cite{Seidel2009NM,Farokhipoor2011PRL}
and photovoltaic\cite{Yang2010NN} properties of ferroelectric DWs
(FE-DWs) have fostered the interest in using them as functional parts
in devices in nanoelectronics.\cite{Catalan2012RMP} The vast majority
of the recent discoveries on FE-DWs have featured multiferroic
BiFeO$_3$ (bismuth ferrite or BFO), a material that has concentrated
many experimental efforts during the past decade because of the
promise it holds for room-temperature applications based on
magnetoelectric effects.\cite{Catalan2009AM} For all these reasons,
currently there is a strong interest in improving our atomistic
understanding of FE-DWs, in particular in the case of BFO.

While the width of magnetic DWs is in the range of 10 to 100~nm,
FE-DWs are believed to be much thinner. The atomic structure of
ferroelectric walls can be studied experimentally using techniques
such as high-resolution transmission electron microscopy, which
permits a spatial definition of up to around 0.5~\AA.  However, the
atomic displacements that characterize FE-DWs are of the order of
0.02~\AA, and therefore direct imaging and interpretation of the
structure of these walls is still a challenge.\cite{Gopalan2007ARMR}
First-principles calculations have also been used to obtain
information about FE-DWs. Some studies\cite{Padilla1996PRB,
Poykko1999APL, Meyer2002PRB, Taherinejad2012PRB} were carried out on
ferroelectric perovskites where the polarization is the only structural order
parameter, such as BaTiO$_3$ and PbTiO$_3$. They report that FE-DWs in
these materials are atomically thin. Other recent first-principles
studies\cite{Seidel2009NM, Lubk2009PRB} have focused on multiferroic
BFO. The width computed for BFO's DWs is at most 1~nm, and some of
them are reported to show a chiral pattern that is abstent in former studies.

In the rhombohedral phase of BFO that is stable at ambient conditions
($R3c$ space group), the O$_{6}$ octahedra of oxygens that surround
the Fe ions are rotated in antiphase about the polarization
axis. These rotations constitute a second structural order parameter
(OP) in addition to the polarization; such an O$_6$-rotation related
OP is usually referred to as {\em anti-ferrodistortive} (AFD). AFD
modes constitute the most usual structural instability in the family
of perovskite oxides, and have been studied in a systematic and
extensive way starting with the classic works of
Glazer\cite{Glazer1972AC} and others. Interestingly, AFD and FE
distortions often compete in perovskite oxides, and it is rare to see
them occur simultaneously in a material.\cite{sichuga11} In fact, as
far as we know, BFO is the only perovskite oxide that presents both
OPs in its equilibrium phase at ambient conditions, which makes it a
unique case for study.

Whenever two OPs occur simultaneously in a given phase of a material,
we tend to assign the lables {\em primary} and {\em secondary} to
them. Such a classification is useful to rationalize the structural
phase transition and domain variants occurring in the crystal: The
primary OP is the one that drives the phase transition and it would
occur even in absence of the secondary OP; the primary OP also defines
the number and properties of the structural domains that can form. In
contrast, the occurrence of the secondary OP relies on the presence of
the primary distortion; the secondary OP may appear as a result of the
symmetry breaking caused by the primary OP (its character would thus
be {\em improper}) or because of a destabilizing coupling with the
primary OP (the secondary order would be {\em triggered} in that
case). Interestingly, as we will see, such a distinction between primary and
secondary OPs cannot be made in BFO's case.

Most of the literature on BFO focuses on the investigation of the FE
polarization, which is effectively treated as the single structural OP
of the compound. This is a rather natural approach: The spontaneous
polarization, with its associated electrostatic energy, is the leading
driving force for the formation of domains; additionally, we can
easily act on it by applying an external field. As regards the AFD
distortions, the oxygen octahedra are known to rotate specifically
about the polarization axis, and are much more difficult to manipulate
and characterize experimentally; hence, they are usually treated as
if they played a secondary role. However, these AFD modes do not occur
as a result of a polarization-induced symmetry breaking; further,
first-principles calculations show that they constitute a genuine structural
instability of the ideal cubic perovskite phase of BFO, independently
of the occurrence of the polarization. Hence, it is clear that the AFD
distortions do not fit the definition of a secondary OP given
above. Moreover, the rhombohedral phase of BFO does not occur as a
consequence of a displacive phase transition characterized by the
condensation of the FE soft mode; rather, BFO's $R3c$ phase undergoes
a strongly discontinuous transformation upon heating, and the
resulting high-temperature structure is probably characterized by
oxygen-octahedra rotations more than it is by polar
distortions. Hence, from this perspective either, there are no clear
arguments to distinguish between primary and secondary OPs in BFO.

Detailed studies based on density functional theory (e.g., see Table~I
of Ref.~\onlinecite{Dieguez2011PRB}) allow us to make this discussion
more quantitative. It has been shown that the FE distortion of the
ideal perovskite phase (which would lower the symmetry from the cubic
$Pm\bar{3}m$ space group to rhombohedral $R3m$) reduces BFO's energy
by about 750~meV/f.u.\ (the precise result depends on the density
functional employed in the calculations~\cite{Dieguez2011PRB}). In
contrast, a pure AFD distortion (leading to the $R\bar{3}c$ space
group) reduces the energy of the cubic phase by about 680~meV/f.u. The
combined FE and AFD distortions render the $R3c$ phase of BFO, which
lies about 925~meV/f.u.\ below the cubic structure. We can thus
conclude that the FE and AFD instabilities compete: If they did not
interact at all, the energy of the phase combining both should be
about 1430~meV/f.u.\ below the cubic structure (with
1430~=~750$+$680); the significantly smaller energy difference
obtained from the simulations (925~meV/f.u.) clearly denotes a
competition.

Hence, the simultaneous occurrence of FE and AFD distortions in BFO's
$R3c$ phase has to be attributed to the fact that, individually
considered, both constitute very strong instabilities of the cubic
phase of the material. Such instabilities clearly compete, not
cooperate; yet, their simultaneous occurrence renders a net reduction
of BFO's energy, thus overcoming the adverse effect of the
competition. Let us note that, because the polarization and rotation
axes coincide, one might be led to believe that the two OPs cooperate
in some way, but such an interpretation would be wrong; rather, the
coincidence of the two axes defines the conditions for which the
FE-AFD competition is weakest.

In view of these facts, one should probably say that BFO presents two
{\em primary} order parameters that need to be considered on equal
footing in a theoretical discussion of this material. Bearing this in
mind, we have conducted a first-principles study to reexamine the
FE-DWs occurring in the $R3c$ phase of bulk BFO. Interestingly, we
find that the DW energetics is essentially determined by the type of
discontinuity affecting the AFD patterns, and not so much by the
change in the FE polarization accross the wall. Our observations have
a rather natural explanation in the context of recent discoveries
highlighting what we may call BFO's {\em polymorphism}, as we are able
to relate the lowest-energy DWs found with the low-lying meta-stable
phases that this compound is known to present. More precisely, the
atomic arrangement {\em at} the most stable FE-DWs resembles the
structure of a different BFO polymorph. Thus, our results suggest that
multi-domain configurations in BFO display a peculiar form of
polytypism. The connection of our work with the novel {\em
  nanoscale-twinned} structures\cite{Prosandeev2012AFM} predicted to
occur in BFO under various contidions (high temperature, high
pressure, appropriate chemical substitution) are also discussed,
offering a coherent picture of structural diversity and energetics in
this material.

The paper is organized as follows. In Section~II we discuss the
details of the configurations we studied, introducing a rigourous
classification of the FE/AFD DWs that can occur in BFO; we also
describe the first-principles methodology used. In Section~III we
present and discuss our results, revealing the structural details that
determine the DW energetics. Other aspects of the DWs, as e.g. their
electronic structure, are also addressed, and the connection of our
results with existing experimental information is discussed. In Section~IV we
summarize the work and present our conclusions.


\section{Methods}


\subsection{Definition of the combined FE/AFD-DWs}

\begin{figure}
\centering
\subfigure[]{
\includegraphics[height=55mm, angle=-90]{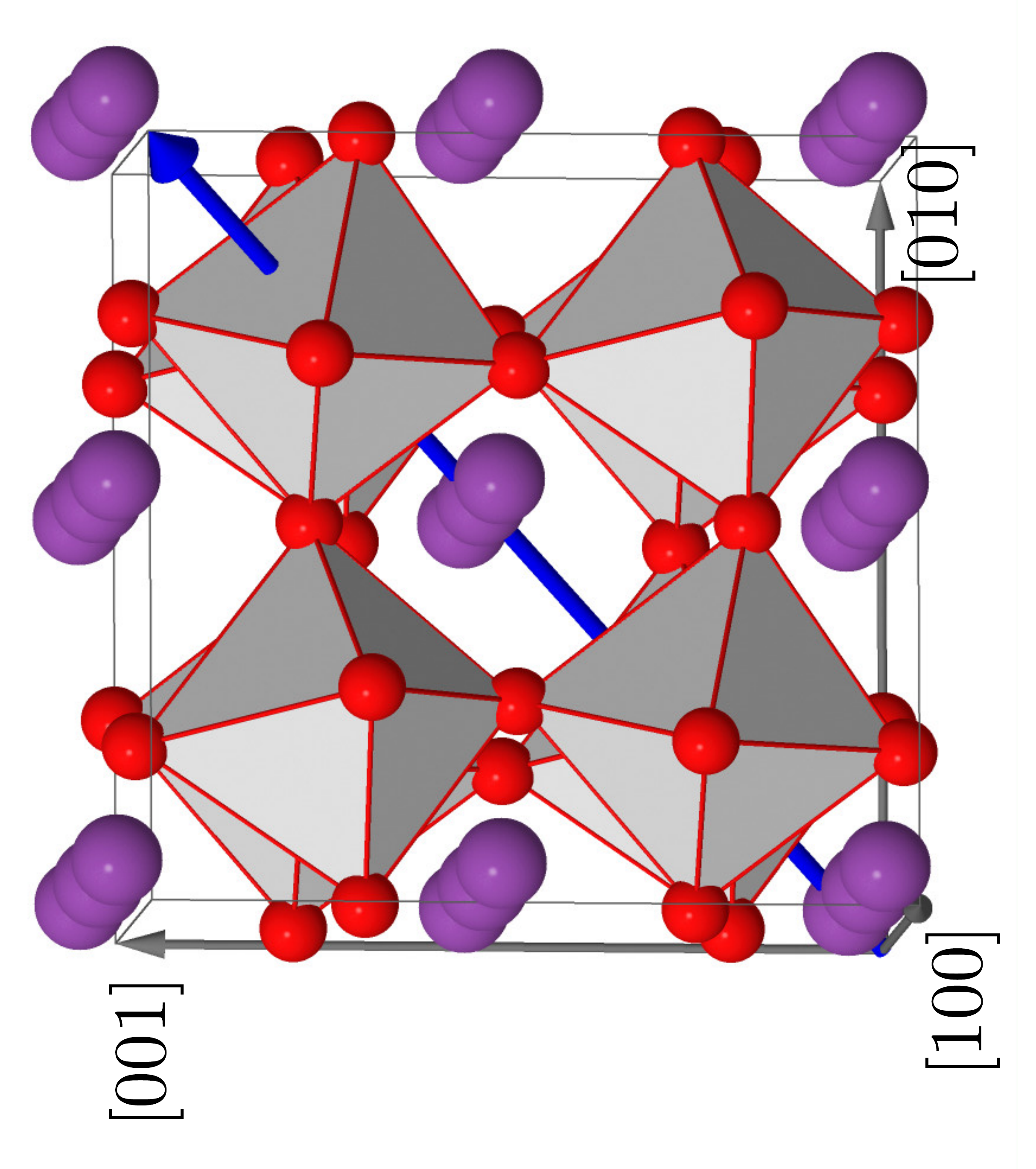}
}
\\
\subfigure[]{
\includegraphics[height=60mm, angle=-90]{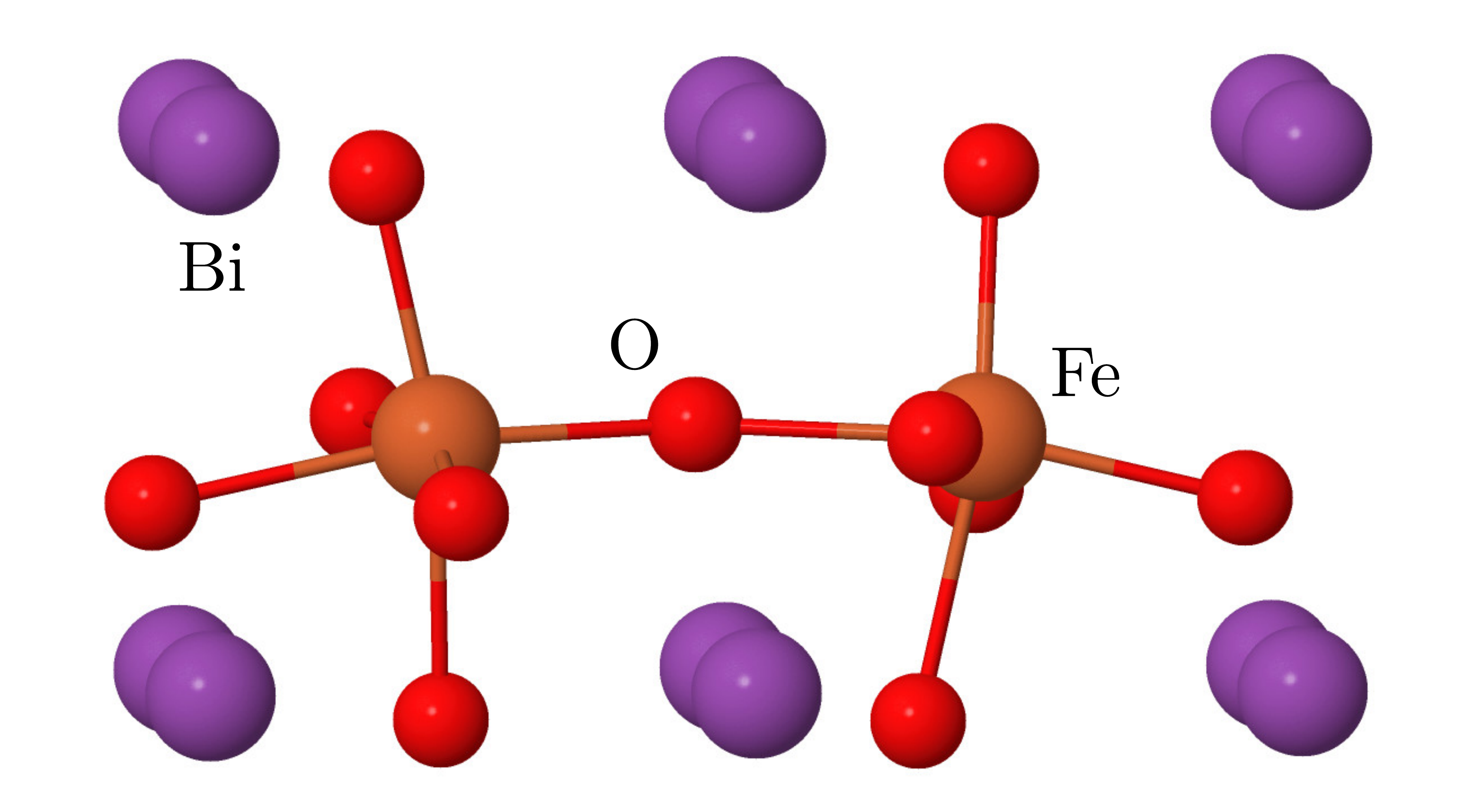}
}
\caption{(Color online.) (a) Pseudocubic cell of BFO showing its three
  main axes (thin arrows) and the polarization along [111]$_{\rm pc}$
  (thick arrow).  (b) Detail of the cell, where two octahedra have
  been removed to allow visualization of Fe cations and Fe-O bonds.  }
\label{fig_pseudocubic}
\end{figure}

BFO crystallizes in rhombohedral space group $R3c$, with a primitive
unit cell that contains 10 atoms.  Its rhombohedral lattice parameters
are $a_{\rm rh} = 5.6343$ \AA~and $\alpha_{\rm rh} =
59.348^\circ$.\cite{Kubel1990AC} It is easier to visualize its
structure using the pseudocubic 40-atom unit cell shown in
Fig.~\ref{fig_pseudocubic}. The Bi and Fe cations are displaced from
their high symmetry positions along the [111]$_{\rm pc}$ pseudocubic
axis, and the material develops a polarization in this direction.  At
the same time, the O$_6$ octahedra surrounding Fe cations rotate about
this [111]$_{\rm pc}$ axis in antiphase, which corresponds to the
so-called $a^{-}a^{-}a^{-}$ pattern in Glazer's
notation.\cite{Glazer1972AC}

Now, the cubic perovskite structure presents 8 equivalent directions
of the $\langle 111 \rangle_{\rm pc}$ type. In a given domain, the
electric polarization $\boldsymbol{P}$ can point along any of these 8
directions; further, once a direction for $\boldsymbol{P}$ is chosen,
there are two equivalent ways (related by a phase shift) in which the
O$_6$ octahedra can rotate. Hence, in total we can have 16 different
FE+AFD domains in the rhombohedral phase of BFO.

We now introduce a notation that will help us to describe the
relationships between these domains. Let $\boldsymbol{\omega}_i$ be
the three-dimensional pseudovector quantifying the rotations of the
O$_6$ octahedron at cell $i$; the components $\omega_{i\alpha}$, with
$\alpha = x, y$, and $z$, pertain to individual rotations about each
of the pseudocubic principal axes of the perovskite structure, which
are sketched in Fig.~\ref{fig_pseudocubic}. (From now on, all vectors
and planes will be referred to the pseudocubic axes unless we mention
otherwise.\cite{fn_reference}) As mentioned earlier, BFO's O$_6$
rotations around $\langle 111 \rangle$ are modulated in antiphase when
we move from one cell to any of its first-nearest-neighboring
cells. Such a modulation corresponds to the $R$ $q$-point of the
Brillouin zone of the ideal cubic 5-atom perovskite cell; for the
particular case of rotations around the [111] axis, the pattern can be
mathematically expressed as $\boldsymbol{\omega}_{i} = \omega_0
(-1)^{n_{ix}+n_{iy}+n_{iz}} (1,1,1)$, where the $n_{i\alpha}$
variables are integers defining the location of cell $i$ in the
lattice.

Let us consider how a FE-DW can affect this ideal pattern. To do that,
let us choose a cell $i$ that is well inside the first domain; let us
assume that $\boldsymbol{\omega}_{i} = \omega_0 (1,1,1)$ and that the
polarization in this first domain is also along [111], so that ${\bf
  P}^{\rm I} = P_0 (1,1,1)$. Then, let us pick a cell $i'$ that is
well inside the second domain, and which can be reached from cell $i$
by advancing an even number of cells along the pseudo-cubic direction
$\alpha$. Thus, we have $n_{i'\beta} - n_{i\beta} =
2n\delta_{\alpha\beta}$, $n$ being an integer. Obviously, the two
domains separated by the DW are related by symmetry, which implies
that $\boldsymbol{\omega}_{i}$ and $\boldsymbol{\omega}_{i'}$ must be
symmetry-related as well. We can have the following possibilities:

(1) $\boldsymbol{\omega}_{i'} = \boldsymbol{\omega}_{i}$, which
coincides with what we would see in the monodomain situation, i.e.,
the pattern of O$_6$ rotations is essentially unaltered by the
FE-DW. We will denote this case as a ``AFD-DW of type 0''. Note that,
since in BFO the O$_6$-rotation axis coincides with the direction of
the spontaneous polarization, we can conclude that the polarization of
the second domain must be either ${\bf P}^{\rm II} = P_0 (1,1,1)$ or
${\bf P}^{\rm II} = P_0 (-1,-1,-1)$; the former possibility implies
there is no FE-DW (``FE-DW of type 0'') and the latter corresponds to
a 180$^{\circ}$ FE-DW (``FE-DW of type 3'', as the three polarization
components change sign). We have thus identified the first two
possible types of FE/AFD-DWs that can occur in BFO, and which we will
denote, respectively, by ``0/0'' (i.e., FE-DW of type~0 and AFD-DW of
type~0, which means that there is no discontinuity at all) and ``3/0''
(i.e., FE-DW of type~3 and AFD-DW of type~0, which means that the
discontinuity only affects the polarization, which rotates by
180$^{\circ}$ when moving from domain~I to domain~II).

(2) One rotation component changes sign with respect to the ideal
situation. For the sake of the discussion, let us assume that the
affected direction is $z$, so that $\boldsymbol{\omega}_{i'} =
\omega_0 (1,1,-1)$. This case, which we will denote as ``AFD-DW of
type 1'', does imply a discontinuity of the $a^{-}a^{-}a^{-}$
rotational pattern. Indeed, we must have some sort of {\em phase
  boundary} for the O$_6$ rotations about the $z$ axis as we cross the
DW. As we will see later, our simulations indicate that such a phase
boundary is the most trivial one that we could imagine: The
$z$-rotations of neighboring cells {\em accross the DW} occur
in-phase, thus breaking the anti-phase modulation sequence of the
ideal pattern. Here again, we can deduce that the polarization of the
second domain must be either ${\bf P}^{\rm II} = P_0 (1,1,-1)$ or
${\bf P}^{\rm II} = P_0 (-1,-1,1)$ implying that the FE-DW must be
either of ``type 1'' (the 71$^{\circ}$ FE-DW in which only one
polarization component changes sign) or of ``type 2'' (the
109$^{\circ}$ FE-DW with two components changing sign),
respectively. We have thus identified two new types of FE/AFD-DWs,
which we denote by ``1/1'' and ``2/1'', respectively.

(3) Two rotation components change sign, so that
e.g. $\boldsymbol{\omega}_{i'} = \omega_0 (1,-1,-1)$. This will be our
``AFD-DW of type 2'', and the associated FE-DW can be either of ``type
2'', with ${\bf P}^{\rm II} = P_0 (1,-1,-1)$, or of ``type 1'', with
${\bf P}^{\rm II} = P_0 (-1,1,1)$. We thus have the combined
FE/AFD-DWs of types ``2/2'' and ``1/2''.

(4) $\boldsymbol{\omega}_{i'} = - \boldsymbol{\omega}_{i}$,
corresponding to a ``AFD-DW of type 3''. In this case, the FE-DW can
either be of ``type 3'', with ${\bf P}^{\rm II} = P_0 (-1,-1,-1)$, or
of ``type 0'' , with ${\bf P}^{\rm II} = P_0 (1,1,1)$. The combined
FE/AFD-DWs are thus denoted by ``3/3'' and ``0/3'',
respectively. Obviously, in the latter case there is no FE-DW, and the
discontinuity only affects the AFD pattern.

\subsection{Domain walls investigated}

In this article we will consider DWs occurring at planes of low Miller
indices, namely, (100) and (110), which are the ones that were
discussed by previous authors.\cite{Seidel2009NM, Lubk2009PRB} Our
initial task is to reduce the number of possible combinations of
domains that we have to study.  First, for a given DW many of these
configurations will be equivalent by symmetry.  Second, we limit
ourselves to {\em neutral} DWs, i.e., those in which there is no
discontinuity in the polarization component across the wall. Note that
such a discontinuity would give raise to bound charges at the DW,
which would lead to a large unfavorable electrostatic
energy.\cite{fn_defects,Wu2006PRB} Hence, the {\em electrostatically
  allowed} cases are those in which $({\bf P}^{\bf I} - {\bf P}^{\rm
  II}) \cdot (1, 0, 0) = 0$ and $({\bf P}^{\bf I} - {\bf P}^{\rm II})
\cdot (1, 1, 0) = 0$, respectively, for the (100) and (110) planes.

Lastly, we could also limit ourselves to domains that fulfill the
condition of {\em mechanical
  compatibility}.\cite{Fousek1969JAP,Streiffer1998JAP} To understand
the origin of this condition, note that 
the ground
state phase of ferroelectric perovskites involves a distortion of the
cubic cell, i.e., an homogeneous strain. In the presence of a DW, the
distortion of unit cells at opposite sides of the DW may or may not
coincide.
If such strains coincide, we have
mechanical compatibility; if not, an elastic energy penalty will
occur. In the case of BFO, a possible mechanical mismatch at the DW
comes from the relatively small shear strains that change the
pseudocubic angles from 90$^{\circ}$ to about 89.2$^{\circ}$. As a
result, in BFO there are pairs of domains that, although do not
exactly fulfill the mechanical compatibility condition, are very close
to doing so. We thus decided to consider them for study.

Once the criteria described in the previous paragraphs are applied, it
turns out that there are 6 possible inequivalent configurations for
(100) DWs, all of them with ${\rm P}^{\rm I} = P_0(1,1,1)$.
Concerning (110) domains, there are two inequivalent ways to fix
domain~I: using ${\bf P}^{\rm I} = P_0(1,1,1)$, or using ${\bf P}^{\rm
  I} = P_0(1,-1,1)$, which implies a domain where the polarization is
parallel to the DW. For these two alternatives we find, respectively,
four and eight inequivalent DW configurations. In total, we have 18
different DWs that we considered in this study. Table \ref{tab_cases}
lists them, indicating their FE/AFD-DW type and whether the mechanical
matching condition is fulfilled or not.

\begin{table}
\caption{Configurations studied. Columns:
(1) polarization direction in domain I,
(2) DW plane,
(3) polarization direction in domain II,
(4) FE/AFD DW-type according to polarization change and octahedra rotation
    pattern change (see text), and
(5) whether the mechanical compatibility condition is met at the DW.
}
\vskip 2mm
\centering
\begin{tabular*}{0.75\columnwidth}{@{\extracolsep{\fill}}ccccc}
\hline\hline
$\bf{P}^{\rm I}$ & DW & $\bf{P}^{\rm II}$     & Type & Mech.? \\
\hline
[111] & (100) & [111]                         & 0/0  &  YES \\[0pt]
[111] & (100) & [111]                         & 0/3  &  YES \\[0pt]
[111] & (100) & [11$\bar{1}$]                 & 1/1  &  NO  \\[0pt]
[111] & (100) & [11$\bar{1}$]                 & 1/2  &  NO  \\[0pt]
[111] & (100) & [1$\bar{1}$$\bar{1}$]         & 2/1  &  YES \\[0pt]
[111] & (100) & [1$\bar{1}$$\bar{1}$]         & 2/2  &  YES \\[0pt]
\hline
[111] & (110) & [111]                         & 0/0  &  YES \\[0pt]
[111] & (110) & [111]                         & 0/3  &  YES \\[0pt]
[111] & (110) & [11$\bar{1}$]                 & 1/1  &  YES \\[0pt]
[111] & (110) & [11$\bar{1}$]                 & 1/2  &  YES \\[0pt]
\hline
[1$\bar{1}$1] & (110) & [1$\bar{1}$1]         & 0/0  &  YES \\[0pt]
[1$\bar{1}$1] & (110) & [1$\bar{1}$1]         & 0/3  &  YES \\[0pt]
[1$\bar{1}$1] & (110) & [1$\bar{1}$$\bar{1}$] & 1/1  &  NO  \\[0pt]
[1$\bar{1}$1] & (110) & [1$\bar{1}$$\bar{1}$] & 1/2  &  NO  \\[0pt]
[1$\bar{1}$1] & (110) & [$\bar{1}$11]         & 2/1  &  NO  \\[0pt]
[1$\bar{1}$1] & (110) & [$\bar{1}$11]         & 2/2  &  NO  \\[0pt]
[1$\bar{1}$1] & (110) & [$\bar{1}$1$\bar{1}$] & 3/0  &  YES \\[0pt]
[1$\bar{1}$1] & (110) & [$\bar{1}$1$\bar{1}$] & 3/3  &  YES \\[0pt]
\hline\hline
\end{tabular*}
\label{tab_cases}
\end{table}


\subsection{First-principles methods}

Our calculations are based on density-functional theory
(DFT).\cite{Hohenberg1964PR} Most of them were done using the
local-density approximation (LDA)\cite{Kohn1965PR, Ceperley1980PRL,
  Perdew1981PRB} for the exchange and correlation functional, although
for comparison we repeated some calculations using two flavors of the
generalized-gradient approximation: the functional of Perdew, Burke,
and Ernzerhof\cite{Perdew1996PRL} (PBE), and its adaptation to
solids\cite{Perdew2008PRL} (PBEsol).  To obtain a better description
of iron's 3$d$ orbitals, we added a Hubbard-$U$
term to the energy of the system, following the prescription of
Dudarev and coworkers.\cite{Dudarev1998PRB}

We used the implementation of this formalism in the {\sc Siesta}
code.\cite{SIESTA} {\sc Siesta} uses a basis set of localized
numerical orbitals, which makes it computationally very efficient.  In
our case, we included in our basis set $s$ (DZ), $p$ (SZ), and $d$
(DZ) orbitals centered at Fe; $s$ (DZ), $p$ (DZ), $d$ (SZ), and $f$
(SZ) orbitals centered at Bi; and $s$ (DZ), $p$ (DZ), and $d$ (SZ)
orbitals centered at O. The SZ and DZ terminology means that we used
either one (single-$\zeta$) or two (double-$\zeta$) radial functions
for each of the angular forms, respectively.\cite{SIESTA} We used
norm-conserving Troullier-Martins
pseudopotentials\cite{Troullier1991PRB} for a efficient treatment of
the interaction between ion cores and valence electrons; the
pseudopotentials were generated in the following configurations: Fe's
$4s^2 3d^6$, Bi's $6s^2 6p^3$, and O's $2s^22p^4$; partial core
corrections were included in all cases. We used a real-space grid to
represent the electronic density, the Hartree and exchange-correlation
potentials, and the matrix elements between basis orbitals; this grid
is equivalent to one generated by a plane-wave basis cut off at
300~Ry.  The integrations in reciprocal space were done using a
Monkhorst-Pack grid equivalent to a $4 \times 4 \times 4$ one for a
5-atom perovskite unit cell.

We also used the {\sc Vasp} code\cite{VASP} for our calculations.
{\sc Vasp} uses plane-waves as basis sets, and therefore allows for a
more systematic numerical treatment of the equations to be solved at
the cost of more computational time.  We used the projector-augmented
wave method\cite{Blochl1994PRB} to represent the interaction of
electrons with ionic cores solving for the following electrons: Fe's
3$s$, 3$p$, 3$d$, and 4$s$; Bi's 5$d$, 6$s$, and 6$p$; and O's 2$s$
and 2$p$.  We generated a plane-wave basis set using a kinetic energy
cutoff of 500~eV.  The integrations in reciprocal space were done
using the same k-point grids as in {\sc Siesta}.

When using {\sc Vasp} we set the Hubbard-$U$ equal to 4~eV, a value
determined by requesting the computed magnetic interactions to be in
quantitative agreement with those obtained from calculations with
hybrid functionals.\cite{hong12} $U$~$\approx$~4~eV has become a
frequent choice in first-principles studies of BFO with this same kind
of formalism, as it leads to qualitatively and semi-quantitatively
correct results for all the properties investigated so far.  Due
mainly to the differences in the treatment of the ion cores, with {\sc
  Siesta} we need to use $U=2$~eV to reproduce the {\sc Vasp}
results for the ground state of BiFeO$_3$; thus, we used this value
for the {\sc Siesta} calculations.

Unless mentioned otherwise, the results in our tables and figures are
the ones obtained with {\sc Vasp}.


\section{Results and discussion}


\subsection{Energetics and DW structure}

\begin{table}
\caption{Energetics of the configurations studied with simulation cells
containing 80 atoms. Column 1: domain type
described following Table \ref{tab_cases}.
Column 2: DW energy according to our simulations;
asterisks indicate configurations where the mechanical matching condition
is not fulfilled.
Column 3: label according to $O_6$ matching pattern (see text). 
}
\vskip 2mm
\centering
\begin{tabular*}{0.9\columnwidth}{@{\extracolsep{\fill}}ccc}
\hline\hline
 DW Type & $E_{\rm DW}$ (mJ/m$^2$) & Matching \\
\hline
[111]  (100)  [111]                          0/0 &   0     & - \\[0pt]
[111]  (100)  [111]                          0/3 & 227     & C \\[0pt]
[111]  (100)  [11$\bar{1}$]                  1/1 & 151$^*$ & B \\[0pt]
[111]  (100)  [11$\bar{1}$]                  1/2 & 147$^*$ & B \\[0pt]
[111]  (100)  [1$\bar{1}$$\bar{1}$]          2/1 &  62     & A \\[0pt]
[111]  (100)  [1$\bar{1}$$\bar{1}$]          2/2 & 319     & C \\[0pt]
\hline
[111]  (110)  [111]                          0/0 &   0     & - \\[0pt]
[111]  (110)  [111]                          0/3 & 254     & C \\[0pt]
[111]  (110)  [11$\bar{1}$]                  1/1 & 152     & B \\[0pt]
[111]  (110)  [11$\bar{1}$]                  1/2 & 178     & B \\[0pt]
\hline
[1$\bar{1}$1]  (110)  [1$\bar{1}$1]          0/0 &   0     & - \\[0pt]
[1$\bar{1}$1]  (110)  [1$\bar{1}$1]          0/3 & 293     & C \\[0pt]
[1$\bar{1}$1]  (110)  [1$\bar{1}$$\bar{1}$]  1/1 & 142$^*$ & B \\[0pt]
[1$\bar{1}$1]  (110)  [1$\bar{1}$$\bar{1}$]  1/2 & 196$^*$ & B \\[0pt]
[1$\bar{1}$1]  (110)  [$\bar{1}$11]          2/1 & 150$^*$ & B \\[0pt]
[1$\bar{1}$1]  (110)  [$\bar{1}$11]          2/2 & 244$^*$ & B \\[0pt]
[1$\bar{1}$1]  (110)  [$\bar{1}$1$\bar{1}$]  3/0 &  74     & A \\[0pt]
[1$\bar{1}$1]  (110)  [$\bar{1}$1$\bar{1}$]  3/3 & 255     & C \\[0pt]
\hline\hline
\end{tabular*}
\label{tab_energies}
\end{table}

As a first step in characterizing our DWs we computed the energy
needed to create them.  In order to do so we designed an 80-atom
periodic simulation cell for each configuration listed in
Table~\ref{tab_cases}, containing two domains of 40 atoms and two DWs.
We then wanted to find the minimum energy configuration for each case
without making any prior assumption on the location of the DW in terms
of parallel ion planes, nor on the specific atomic arrangement
occurring at the DW. Thus, for each DW type considered, we started
with some reasonable atomic configuration corresponding to the two
domains involved, and performed a short molecular dynamics simulation,
starting from random velocities, using {\sc Siesta}. Our tests
indicated that this procedure leads to an unrestricted and efficient
exploration of the energy surface, allowing the system to find the
lowest-energy DW configuration in a reliable and reproducible way. We
then relaxed the lattice vectors and atomic positions until the
stresses and forces were small. The configurations thus obtained were
further relaxed using {\sc Vasp} until the forces were below
0.01~eV/\AA~and the stresses were below 0.1~kbar. At all times the
magnetic moments of the Fe ions kept an antiferromagnetic arrangement
of type~G (closest neighbors have antiparallel spins); this is the
configuration that BFO shows experimentally at a local scale at
ambient conditions. Table~\ref{tab_energies} includes all the DW
energies that resulted after this process, computed as
\begin{equation}
\label{eq_eDW}
E_{\rm DW} = \frac{E-E_0}{2S}
\end{equation}
where $E$ is the energy of the DW configuration, $E_0$ is the energy
of bulk BFO (computed for the same supercell for better numerical
accuracy), $S$ is the surface area of the cell face parallel to the
DW, and we divide by two because each cell contains two DWs.

Before discussing our results, let us note that the $E_{\rm DW}$
values that we obtain are significantly lower that those reported in
the previous studies of Seidel {\em et al.}\cite{Seidel2009NM} and
Lubk {\em et al.}\cite{Lubk2009PRB} A comparison is shown in
Table~\ref{tab_lubk_siesta}, where we include our results for the
lowest-energy DWs that involve polarization rotations by 71$^{\circ}$,
109$^{\circ}$, and 180$^{\circ}$, respectively.  For consistency with
the previous studies, all calculations reported in this Table were
done with cells that contained 120 atoms, and we have restricted
ourselves to cases satisfying the mechanical compatibility
conditions. (Our DW energies obtained using 120-atom cells are very
similar to the ones we calculated with 80-atom cells, implying that
the results of Table~\ref{tab_energies} are well converged in this
regard.) There are some minor differences between the {\sc
  Vasp} calculations done by the authors of
Refs.~\onlinecite{Seidel2009NM} and \onlinecite{Lubk2009PRB} and ours; yet,
we have tested that even if we modify our numerical approximations to
make them as similar as possible to theirs, our results change very
little. Moreover, Table~\ref{tab_lubk_siesta} also includes the
results we obtaind with {\sc Siesta}, which are very similar to the
values found with {\sc VASP}. In addition, let us note that similarly
low DW energies have been obtained by Wei {\sl et
  al}.\cite{wei-unpublished} in an investigation of FE-DWs in BFO thin
films under various epitaxial strain conditions. What can explain the
discrepancies with the previous published literature? As some of us
have recently shown,\cite{Dieguez2011PRB} the energy surface of BFO
presents a very large number of local minima, which seems to be mainly
a consequence of the many bonding complexes allowed by Bi's chemical
versatility. We reached exactly the same conclusions in the present
investigation of BFO's DWs, as we found that many possible DW
configurations can render a local minimum of the energy
surface. Naturally, the physical relevance of high-lying local energy
minima is questionable, and we found ourselves forced to implement a
careful search for the {\em global} energy minimum for each DW
type. Indeed, by thermally agitating and then quenching our
configurations during the optimization process, we have had access to
minima with (much) lower energy than the ones previously
reported. Hence, our results stand a better chance of being physically
relevant.

\begin{table}
\caption{Energetics of the configurations with minimum energy for each type
of FE-DW that fulfills the mechanical matching condition
(simulation cells with 120 atoms):
[111](100)[1$\bar{1}$$\bar{1}$] 2/1 ($109^\circ$),
[1$\bar{1}$1](110)[$\bar{1}$1$\bar{1}$] 3/0 ($180^\circ$), and
[111](110)[11$\bar{1}$] 1/1 ($71^\circ$).
}
\vskip 2mm
\centering
\begin{tabular*}{1.0\columnwidth}{@{\extracolsep{\fill}}cccc}
\hline\hline
                         & \multicolumn{3}{c}{$E_{\rm DW}$ (mJ/m$^2$)} \\
         & This work & This work & Lubk {\em et al.}\cite{Lubk2009PRB} \\
 DW Type & ({\sf Siesta}) & ({\sf Vasp}) & ({\sf Vasp}) \\
\hline
 109$^\circ$ &  63 &  62 & 205 \\
 180$^\circ$ &  86 &  82 & 829 \\
  71$^\circ$ & 197 & 167 & 363 \\
\hline \hline
\end{tabular*}
\label{tab_lubk_siesta}
\end{table}

\begin{figure*}
\centering
\subfigure[]{
\includegraphics[width=30mm, angle=-90]{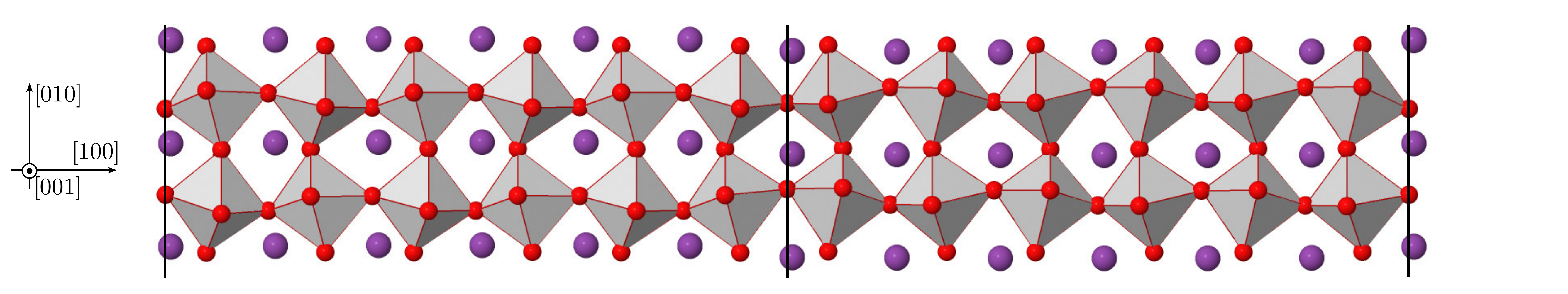}
}
\\
\subfigure[]{
\includegraphics[width=32mm, angle=-90]{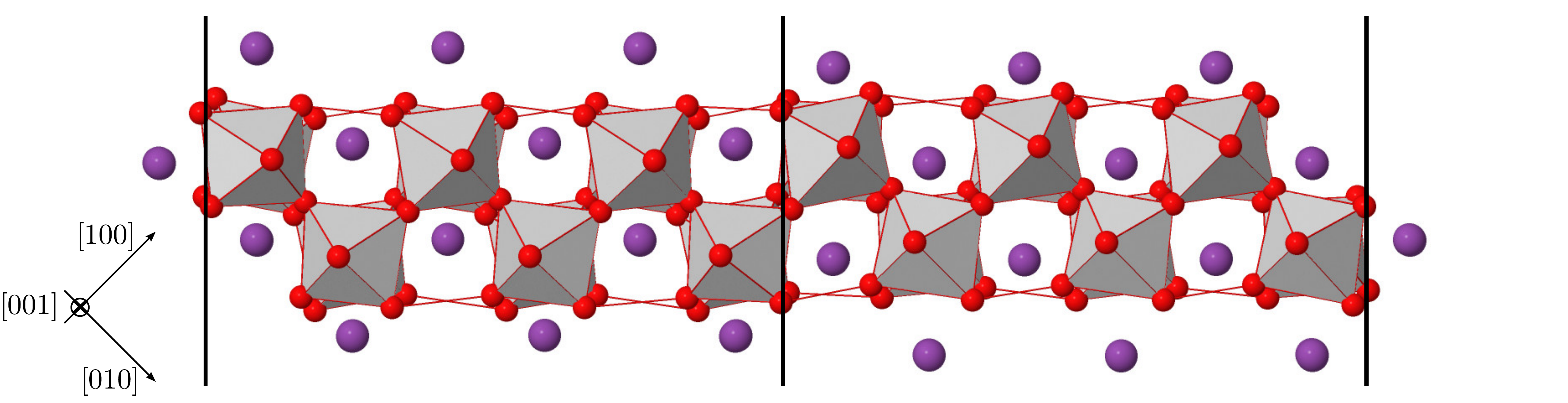}
}
\\
\subfigure[]{
\includegraphics[width=33mm, angle=-90]{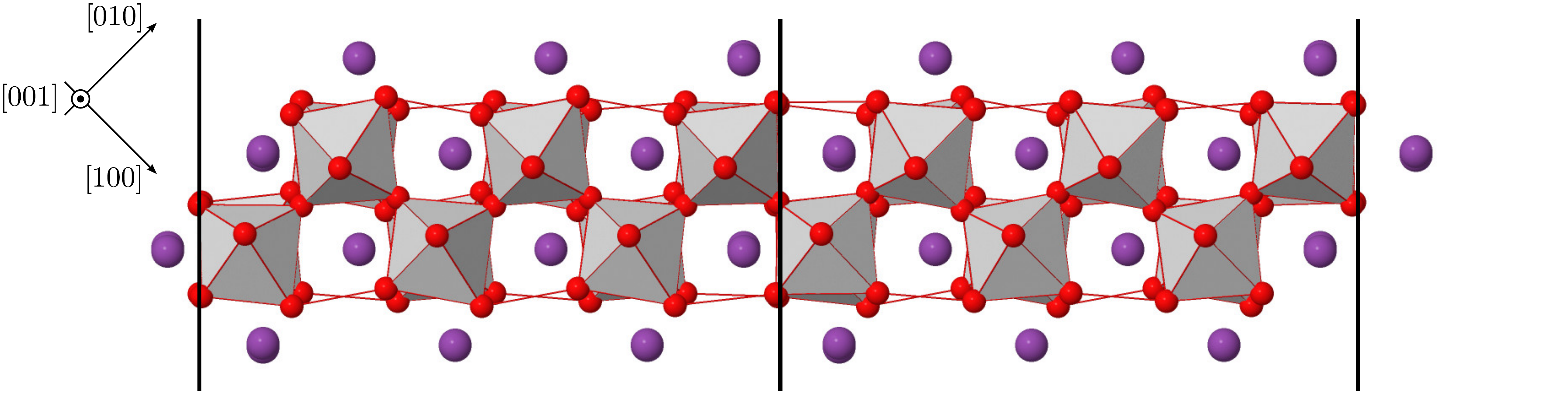}
}
\caption{(Color online.) 
Atomic structures obtained upon relaxation of 120-atom unit cells:
(a) [111](100)[1$\bar{1}$$\bar{1}$] 2/1 DW configuration;
(b) [1$\bar{1}$1](110)[$\bar{1}$1$\bar{1}$] 3/0 configuration;
(c) [111](110)[11$\bar{1}$] 1/1 configuration.
The vertical lines indicate the approximate position of the center of the DWs.
}
\label{fig_dwatoms}
\end{figure*}

\begin{figure}
\centering
\includegraphics[width=85mm]{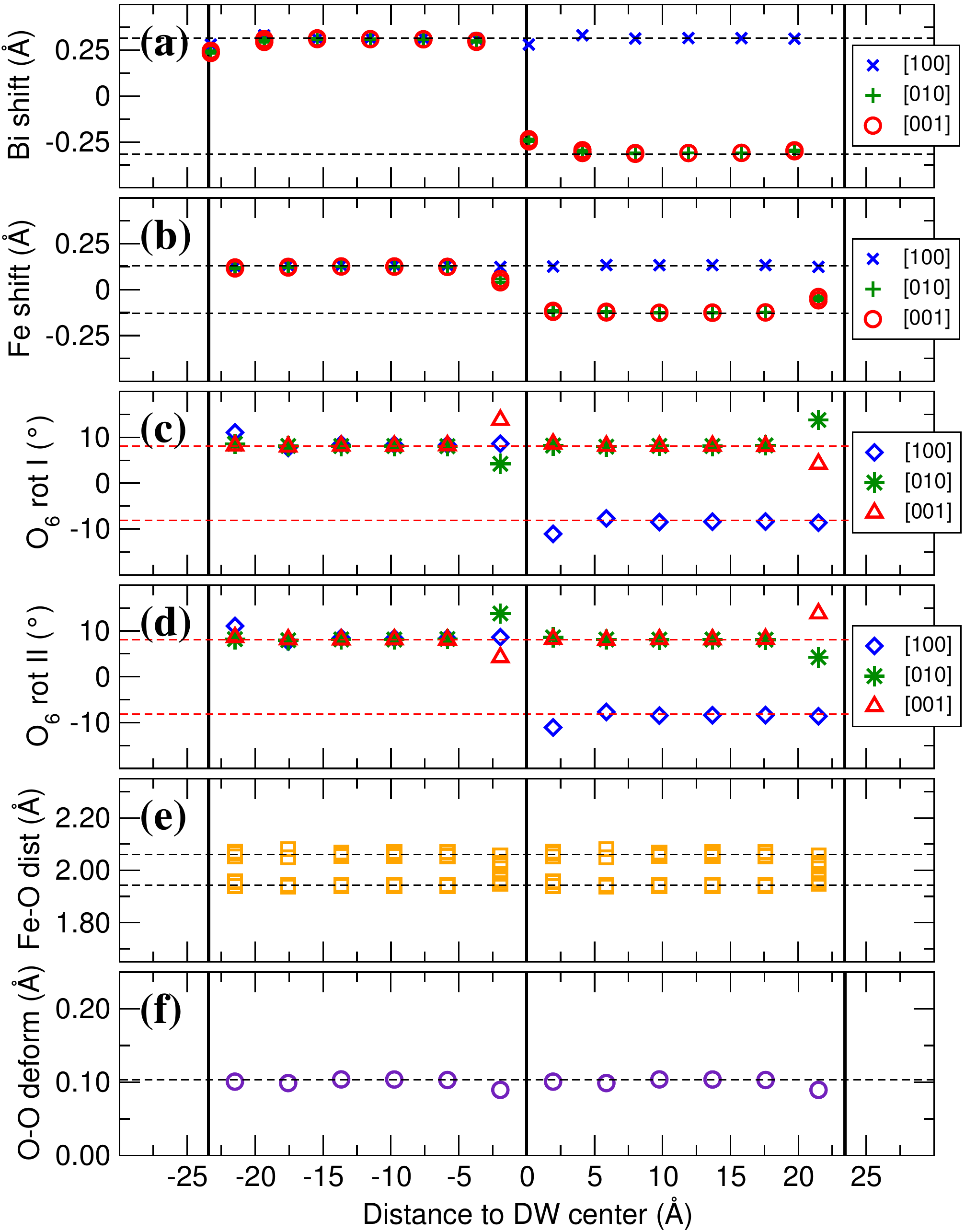}
\caption{(Color online.)  Values of magnitudes that characterize the
  atomic displacements in the [$111$]($100$)[$1\bar{1}\bar{1}$] 2/1 DW
  configuration (polarizations at an angle of around 109$^\circ$).
  The discontinuous lines show the values computed for bulk BFO.  The
  continuous vertical lines mark the approximate position of the
  center of the DW.  We describe in the text those magnitudes that are
  not obvious from the vertical axis labels.  }
\label{fig_dw5}
\end{figure}

We found that the [111](100)[1$\bar{1}$$\bar{1}$]~2/1~DW configuration
has the lowest DW energy. It occurs when two polarization components
and one rotation component change sign at the DW (this is a
109$^{\circ}$~DW from the perspective of the electric
polarization). The relaxed atomic structure that we obtained is shown
in Fig.~\ref{fig_dwatoms}(a); here one can already appreciate that the
DW discontinuity is very sharp, and the O$_6$ octahedra are not
strongly distorted at or near the DW plane. Figures \ref{fig_dw5}(a)
and \ref{fig_dw5}(b) contain the values of the displacements of Bi and
Fe cations along the pseudocubic axes, relative to the center of mass
of the corresponding surrounding O$_{12}$ dodecahedron and O$_{6}$
octahedron, respectively. Such distortions are the origin of the local
polar dipoles giving raise to BFO's macroscopic polarization, and thus
allow us to monitor the $\bf{P}$-change at the DW. The Fe/Bi
displacements are plotted as a function of the distance between the
corresponding cations and the center of the DW, which is assumed to be
located in between the two planes of Fe ions whose environment is most
different from the one they see in bulk BFO. We can thus see that
these polar displacements are almost identical to the ones in bulk,
except for small distortions in a very narrow area around the DW
center. The O$_6$ rotations are quantified in Figs.~\ref{fig_dw5}(c)
and \ref{fig_dw5}(d); they are the three components of
$(-1)^{n_{ix}+n_{iy}+n_{iz}} \boldsymbol{\omega}_{i}$, in the notation
of Section II.A, with the two graphs containing information for the
two inequivalent lines of neighboring O$_6$ octahedra that run
towards the DW.  The O$_6$ rotation patterns are also very
similar to the bulk ones, except for the discontinuity of one AFD
component at the DW.  Concerning Fe-O distances, Fig.~\ref{fig_dw5}(e)
shows again minor distortions confined to the DW.  Finally, in
Fig.~\ref{fig_dw5}(f) we quantify the distortions of the 12 O-O bonds
that form the edges of each O$_6$ octahedron; we do so by computing
the root mean square of the differences between each of the O-O bond
lengths in each O$_6$ octahedron, $d_j$, and their average length
$\bar{d}$,
\begin{equation}
d^{\rm rms} = \sqrt{ \frac{1}{12} \sum_{j=1}^{12} 
\left( d_i - \bar{d} \right)^2} \, .
\end{equation}
This is zero for a regular octahedron, but takes a value of about 0.1
in bulk BFO, as the polar displacements cause variations in the
lengths of the O$_6$ edges. We see that this measure shows very small
differences in O$_6$ octahedra distortions with respect to the bulk
ones. In all, the result of the detailed analysis of the atomic
structure points to a narrow DW whose thickness is around 1~nm.

Our finding that the most stable DW configuration is precisely
[111](100)[1$\bar{1}$$\bar{1}$]~2/1 might seem somewhat arbitrary; on
the contrary, this result leads to very suggestive conclusions. In
this DW, the AFD component that changes sign is the one perpendicular
to the wall; thus, if we focus on the two planes of octahedra next to
the DW, we see that they display a $a^{+}b^{-}b^{-}$ Glazer rotation
pattern, i.e., the O$_6$ tilts occur in phase ($+$ superscript) about
the [100] direction perpendicular to the DW and in antiphase ($-$
superscript) about the [010] and [001] directions within the DW
plane. Further, the discontinuity in the FE polarization -- which is
mainly captured by the Bi displacements shown in Fig.~\ref{fig_dw5}(a)
-- occurs in the plane of the DW, which results in an anti-polar
pattern of Bi displacements around the DW center.  Interestingly, such
structural features are exactly the ones characterizing the $Pnma$
phase of bulk BFO, which is experimentally known to occur at high
temperatures\cite{Arnold2010AFM} and under hydrostatic
pressure,\cite{Haumont2009PRB,Guennou2011PRB} and which is relatively
close in energy to the $R3c$ ground state according to previous
first-principles calculations.\cite{Dieguez2011PRB} In fact, BFO's
$Pnma$ phase would match the structure of our DW almost identically if
we added to it a polar distortion along the direction of the in-phase
oxygen-octahedra rotations. Interestingly, first-principles theory
suggests that such a ferroelectrically-distorted $Pnma$ phase, which
would present the polar space group $Pna2_1$, may be a low-lying
meta-stable polymorph of
BFO,\cite{Dieguez2011PRB,GonzalezVazquez2012PRB} although such a
structure has not been observed experimentally yet. Hence, our results
suggest that the lowest-energy DW found in BFO owes its stability to
the fact that it can adopt an atomic arrangement that mimicks a
low-energy polymorph of the bulk material.

Interestingly, our results for this DW are also reminiscent of the
so-called {\em nanoscale-twinned} phases that have been predicted to
occur in BFO under various conditions (e.g., high temperature,
hydrostatic pressure, chemical substitution of Bi by rare-earth
cations).\cite{Prosandeev2012AFM} In fact, the novel phases described
by Prosandeev {\sl et al}. in Ref.~\onlinecite{Prosandeev2012AFM},
which are claimed to act as a structural bridge between the $R3c$ and
$Pnma$ structures that appear in BFO's phase diagram, can be viewed as
a sequence, along the [100] direction, of DWs of the type that we have
just discussed. It is thus conceivable that, as we heat up BFO's $R3c$
phase, the $Pnma$ structure will nucleate at DWs as the ones just
described, giving raise to intermediate bridging phases of a polytypic
nature.

\begin{figure}
\centering
\includegraphics[width=85mm]{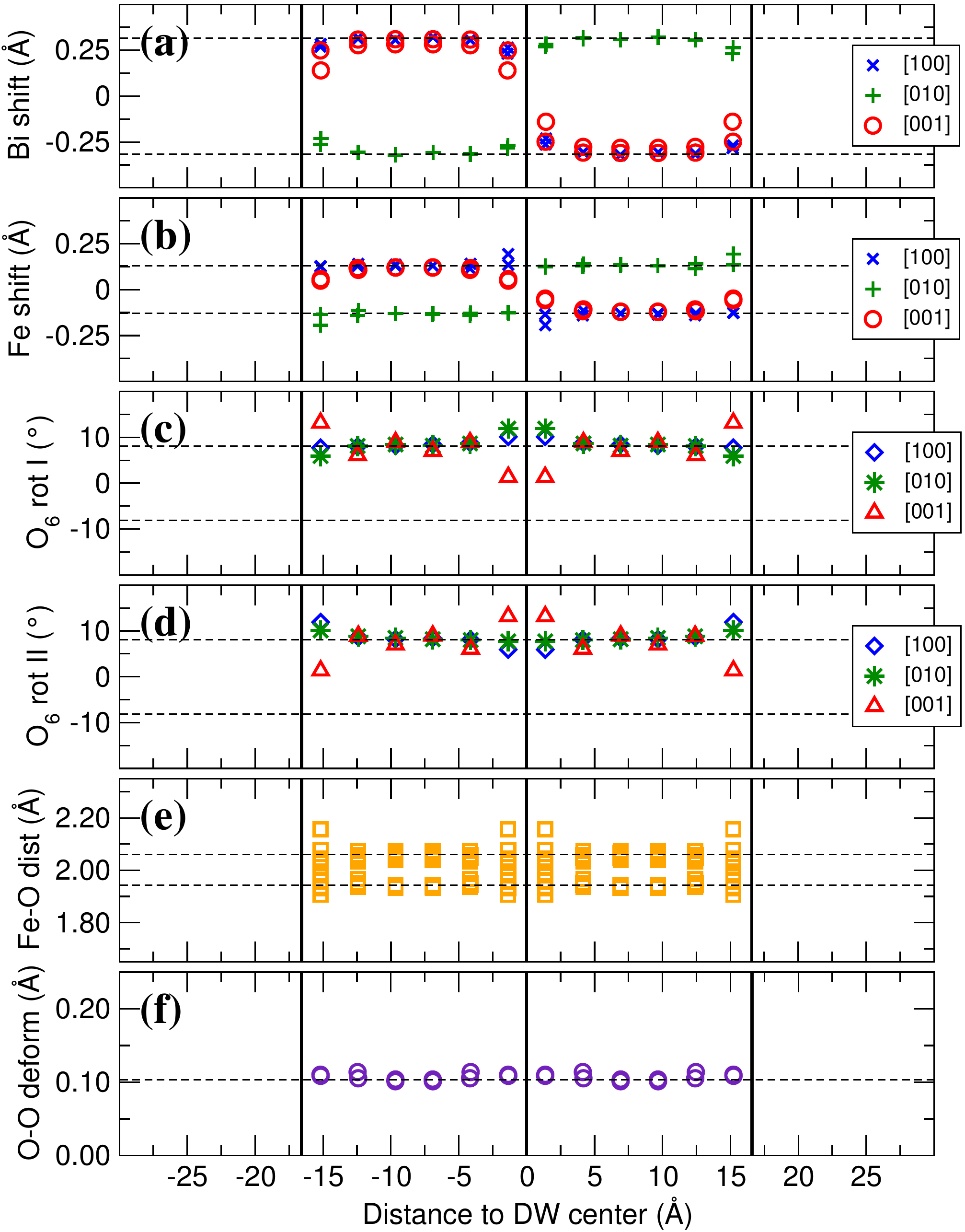}
\caption{(Color online.)  Same measures as in Fig.~\ref{fig_dw5}, but
  for the [1$\bar{1}$1](110)[$\bar{1}$1$\bar{1}$] 3/0 DW configuration
  (polarizations at an angle of 180$^\circ$).  }
\label{fig_dw17}
\end{figure}

The second DW configuration that we found to display a similarly low
DW energy is the [1$\bar{1}$1](110)[$\bar{1}$1$\bar{1}$]~3/0~DW, which
is sketched in Fig.~\ref{fig_dwatoms}(b).  This is a 180$^\circ$~FE-DW
where the cation displacements go in opposite directions at opposite
sides of the wall, as shown in Figs.~\ref{fig_dw17}(a) and
\ref{fig_dw17}(b).  As regards the network of O$_6$ octahedra, there
is no AFD discontinuity involved in this boundary; yet, as we can see
in Figs.~\ref{fig_dw17}(c) and \ref{fig_dw17}(d), because the lattice
needs to accommodate the large change in the polar cation
displacements, the O$_6$ rotation angles at the DW deviate noticeably
from the bulk-like values. From Fig.~\ref{fig_dw17}(e) we notice that
there are some Fe-O bond lenght variations accompanying these
distortions. However, the O-O bond length dispersion is very similar
to the bulk one, as illustrated in Fig.~\ref{fig_dw17}(f).  Let us
also note that Lubk {\em et al.} computed a very high energy of
829~mJ/m$^2$ for their most stable 180$^\circ$~FE-DW which, as in our
case, involves no AFD phase boundary; they also reported that the
polarization rotates in two steps from one domain to the other,
forming a sort of chiral pattern. Our finding of a sharp
180$^\circ$~FE-DW with an energy of 82 mJ/m$^2$ and around 1~nm in
thickness indicates that the configuration described by Lubk {\sl et
  al}. is very unlikely to occur in reality.

\begin{figure}
\centering
\includegraphics[width=85mm]{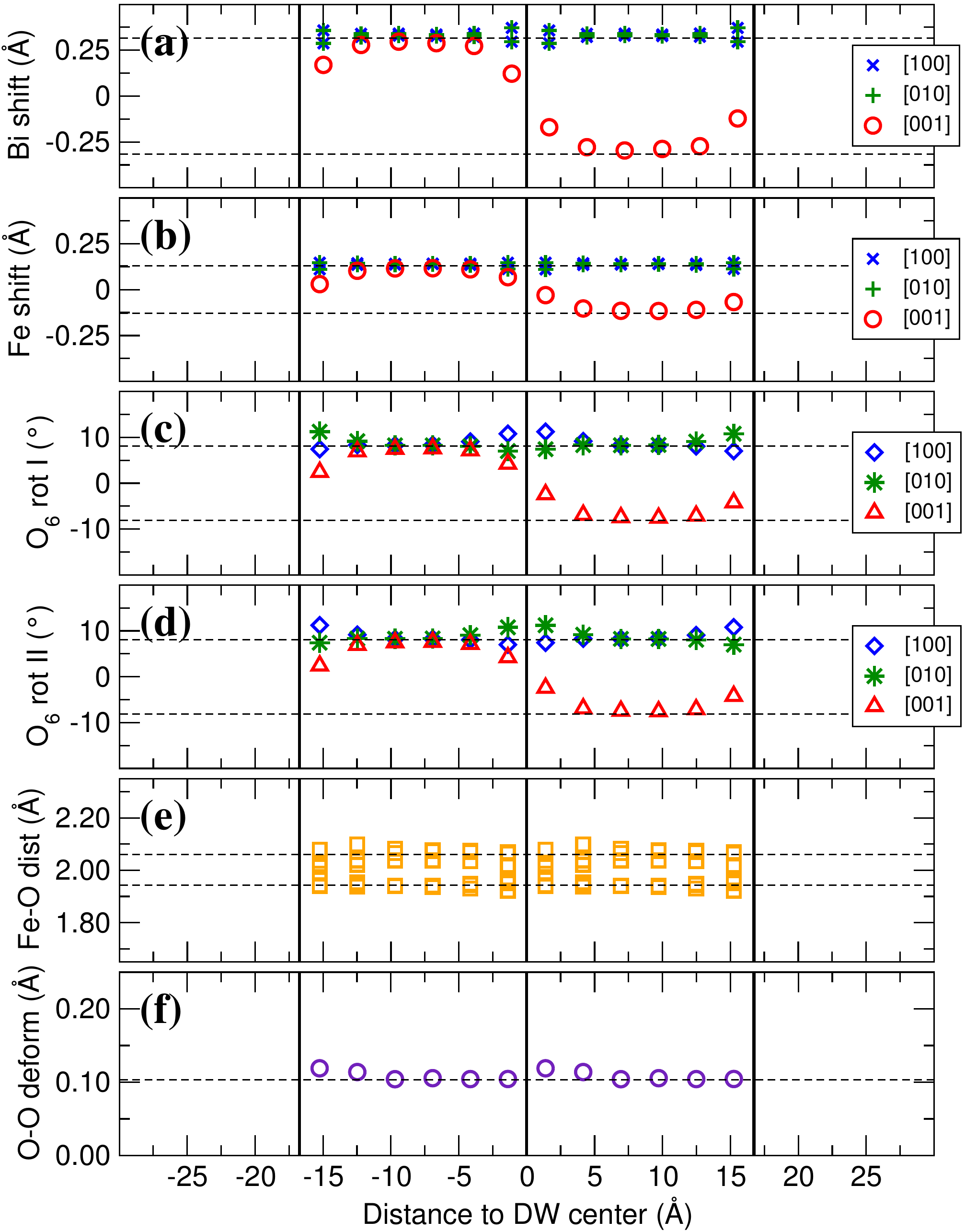}
\caption{(Color online.)  Same measures as in Fig.~\ref{fig_dw5}, but
  for the [111](110)[11$\bar{1}$] 1/1 DW configuration (polarizations
  at an angle of around 71$^\circ$).  }
\label{fig_dw4}
\end{figure}

\begin{figure}
\centering
\includegraphics[width=85mm]{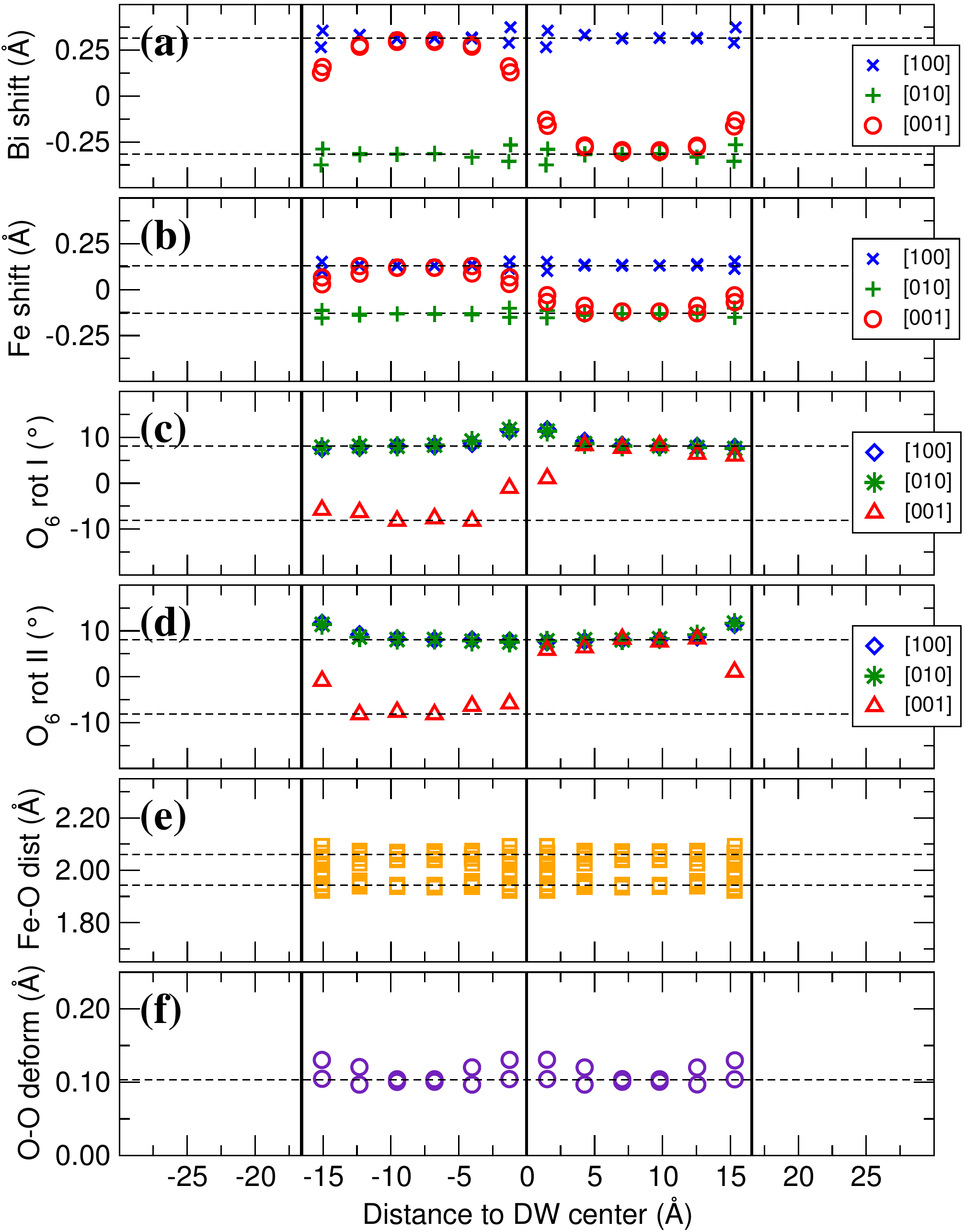}
\caption{(Color online.)  Same measures as in Fig.~\ref{fig_dw5}, but
  for the [1$\bar{1}$1](110)[1$\bar{1}$$\bar{1}$] 1/1 DW configuration
  (polarizations
  at an angle of around 71$^\circ$; matching condition not fulfilled). }
\label{fig_dwN}
\end{figure}

Among all the considered DW configurations, let us discuss in some
detail two additional cases, namely, the [111](110)[11$\bar{1}$] 1/1~DW
that is sketched in
Fig.~\ref{fig_dwatoms}(c) and whose structural features are shown in
Fig.~\ref{fig_dw4}, and the [1$\bar{1}$1](110)[1$\bar{1}$$\bar{1}$] 1/1~DW
characterized by the results in
Fig.~\ref{fig_dwN}. Both of them are 71$^{\circ}$~FE-DWs, their
respective energies being 152~mJ/m$^2$ and
142~mJ/m$^2$ for the 80-atom unit cell.
Interestingly, while the former satisfies the
mechanical compatibility condition mentioned
above,\cite{Fousek1969JAP,Streiffer1998JAP} the latter does not. Yet,
in spite of this fact, the second DW has the lower energy. Indeed,
generally speaking, our results show that the mechanical mismatch does
not involve a large energy penalty. The relative small importance of
this mechanical mismatch should probably be attributed to the small
shear strains associated with the BFO's $R3c$ phase, which result in a
small deformation of the DW plane of little energetic
consequence. Hence, one should probably not consider the
mechanical-matching criterion in studies of rhombohedral BFO, and
should be cautious about its assumption for other materials or phases
displaying small shears.

\begin{figure}[t]
\centering
\includegraphics[height=85mm, angle=-90]{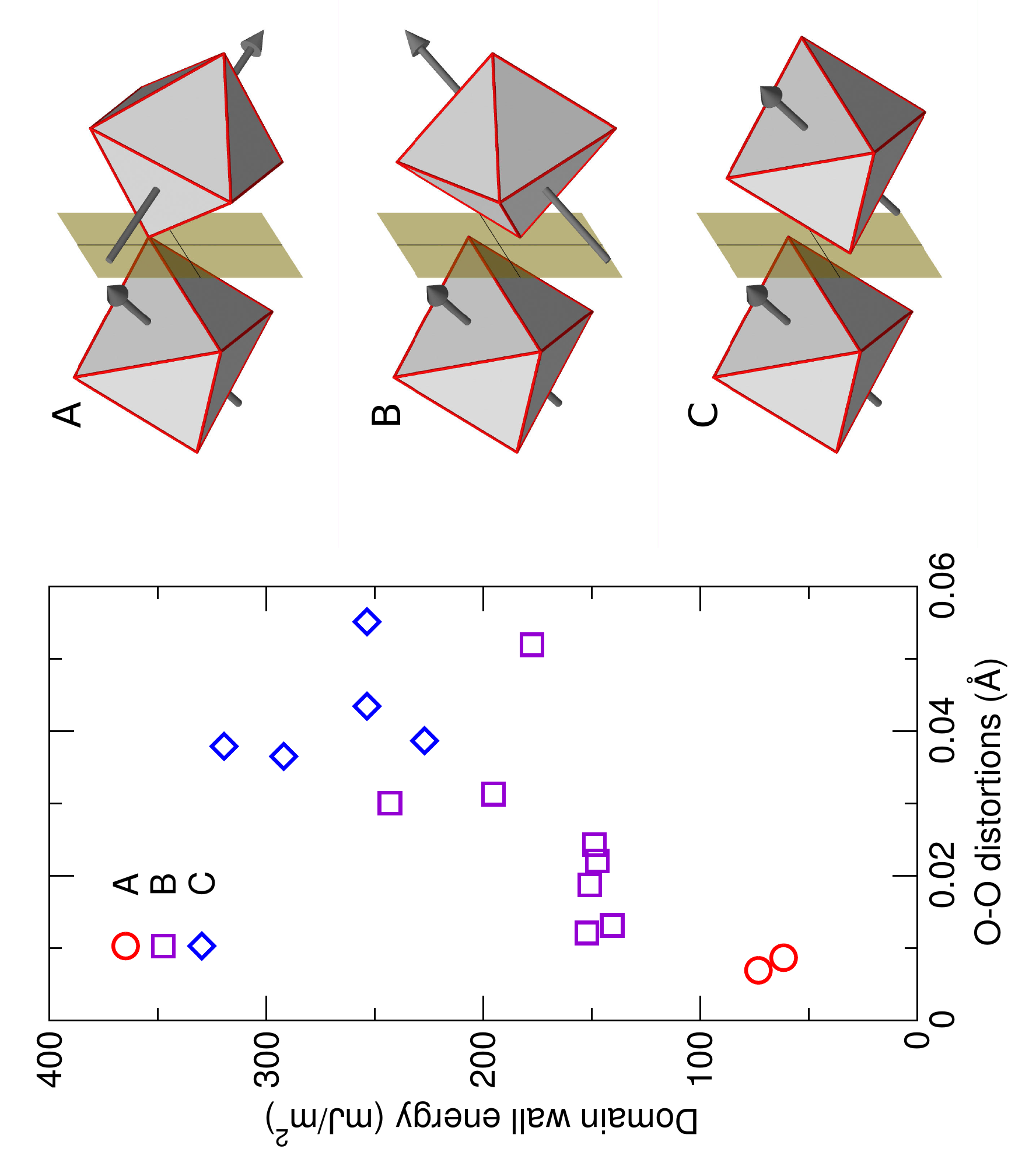}
\caption{(Color online.)  Left panel: DW energy as a function of a
  measure of distortions in O-O edges of O$_6$ octahedra (see text);
  the symbols refer to different types of octahedra mismatching at the
  DW.  Right panel: Illustration of octahedra mismatches, as described
  in the text; each arrow indicates the axis around which each
  octahedron rotates.}
\label{fig_o6dist}
\end{figure}

From the three cases studied in detail so far, we see that low DW
energies are associated to configurations in which the O-O bonds keep
lengths similar to the ones displayed in bulk BFO.  To further
investigate this trend, we quantified the O-O distortion by measuring
the dispersion $d^{\rm rms}_i$ of the $N$ points in graphs such as
that in Fig.~\ref{fig_dw5}(f), relative to the values in bulk BFO,
$d^{\rm rms}_0$,
\begin{equation}
D^{\rm rms} = \sqrt{ \frac{1}{N} \sum_{i=1}^{N} 
\left( d^{\rm rms}_i - d^{\rm rms}_0 \right)^2} .
\end{equation}
We computed this measure for each of the DW configurations of
Table~\ref{tab_energies}, and plotted the DW energy as a function of
it in Fig.~\ref{fig_o6dist}.  We see a very clear correlation between
low DW energies and low variations in the O-O bond lengths.

We can further understand this trend by looking at the rotations that
are present in the center of the domains, independently of the exact
atomic rearrangements that take place at the DW.  Let us consider for
a moment a cubic perovskite crystal where we remove the A and B
cations, and we keep the network of regular
O$_6$ octahedra.  We now impose the formation of an AFD-DW, so that at each
side of it the octahedra rotate a fixed angle, in $a^{-}a^{-}a^{-}$
fashion, about one of the $\langle 111 \rangle$ axes. The oxygen atoms
at the DW plane are shared by two 
octahedra, one from each of the domains separated by the wall. Then,
when the two different $a^{-}a^{-}a^{-}$-like patterns freeze in, the rotation
of each octahedra will {\em try} to send the oxygen atoms
moving in one of four possible directions that form 90$^\circ$ among
them. Depending on which directions are {\em chosen} by the two
octahedra sharing 
a single oxygen, three different cases can arise, as depicted 
in Fig.~\ref{fig_o6dist}: the two directions may coincide
(case~A), they may form 90$^\circ$ (case~B), or they may form
180$^\circ$ (case~C).  In order to accommodate the two conflicting
movements required from a single oxygen ion, the octahedra at the DW
will distort from their regular shape in the 90$^\circ$~case, and more
so in the 180$^\circ$~case.

In a material where AFD and FE distortions are present, the O$_6$
octahedra are more distorted than in the case we have just described.
However, if the rotation angle is relatively large, as in BFO, most of
the displacement of the O atoms out of high-simmetry positions is due
to the rotation about $\langle 111 \rangle$.  We can therefore apply
the analysis of the previous paragraph to our BFO DWs, and label them
following the same notation, since the kind of mismatch is the same
for every pair of octahedra on opposite sides of the wall. We have
done so in Table~\ref{tab_energies}. As expected, we see that type~A
DWs are the lowest in energy, followed by type~B DWs; finally, we find
that type~C DWs tend to be the most unfavorable ones. This is also
illustrated in Fig.~\ref{fig_o6dist}.


\subsection{Role of exchange-correlation functionals}

In a previous article\cite{Dieguez2011PRB} some of us pointed out how
the use of different approximations to the exchange-correlation
functional of DFT can affect the results of the total energy of BFO
configurations.  In particular, we found that the energy differences
between phases (i.e., their relative stability) depends significantly
on the employed energy functional. Therefore, since we may in
principle expect a similarly important dependence of the calculated DW
energies, we repeated our calculations for three representative DW
configurations using two other functionals: PBE\cite{Perdew1996PRL}
and PBEsol.\cite{Perdew2008PRL} Table \ref{tab_xc} shows small
variations in the numbers obtained for the DW energies, while the
relative ordering of the configurations is always the same.  The
variations are smaller than those in Ref.~\onlinecite{Dieguez2011PRB}
because our DW 
configurations are all relatively similar, while in the cited work
phases with very different structures (including, e.g., supertetragonal phases
with O$_5$ pyramids instead of O$_6$ octahedra) were compared.  We take these
results as further confirmation that DFT predicts that the most stable
DW configurations are those that keep the O$_6$ octahedra bond
distortions to a minimum, and that their DW energies are in the range
of several tens of mJ/m$^2$.

\begin{table}[b]
\caption{Energies (in mJ/m$^2$) of the three representative DW
  configurations of Table~\ref{tab_lubk_siesta}, when studied with
  three different approximations for the exchange-correlation
  functional (simulation cells with 80 atoms).  }
\vskip 2mm
\centering
\begin{tabular}{cccc}
\hline\hline
DW Type  &  LDA+$U$  &  PBE+$U$ &  PBEsol+$U$  \\
\hline
109$^\circ$    &  62   &  78  &  59  \\[0pt]
180$^\circ$    &  73   &  86  &  94  \\[0pt]
 71$^\circ$    & 152   & 136  & 146  \\[0pt]
\hline\hline
\end{tabular}
\label{tab_xc}
\end{table}


\subsection{Electronic structure}

Seidel and collaborators\cite{Seidel2009NM} were the first to report
the experimental observation of an enhanced electronic conductivity at
room temperature in BFO DWs.  In their article, they also included
results about the electronic properties of the DWs obtained using
first-principles calculations; they mentioned that at the center of
the domains the local density of states (LDoS) resembles the one of
the bulk, while as the DW is approached the bandgap is reduced by an
amount of up to 0.2~eV depending on the DW orientation.  They
correlated this reduction with the presence of the conductivity at the
walls.

\begin{figure}
\centering
\includegraphics[width=75mm]{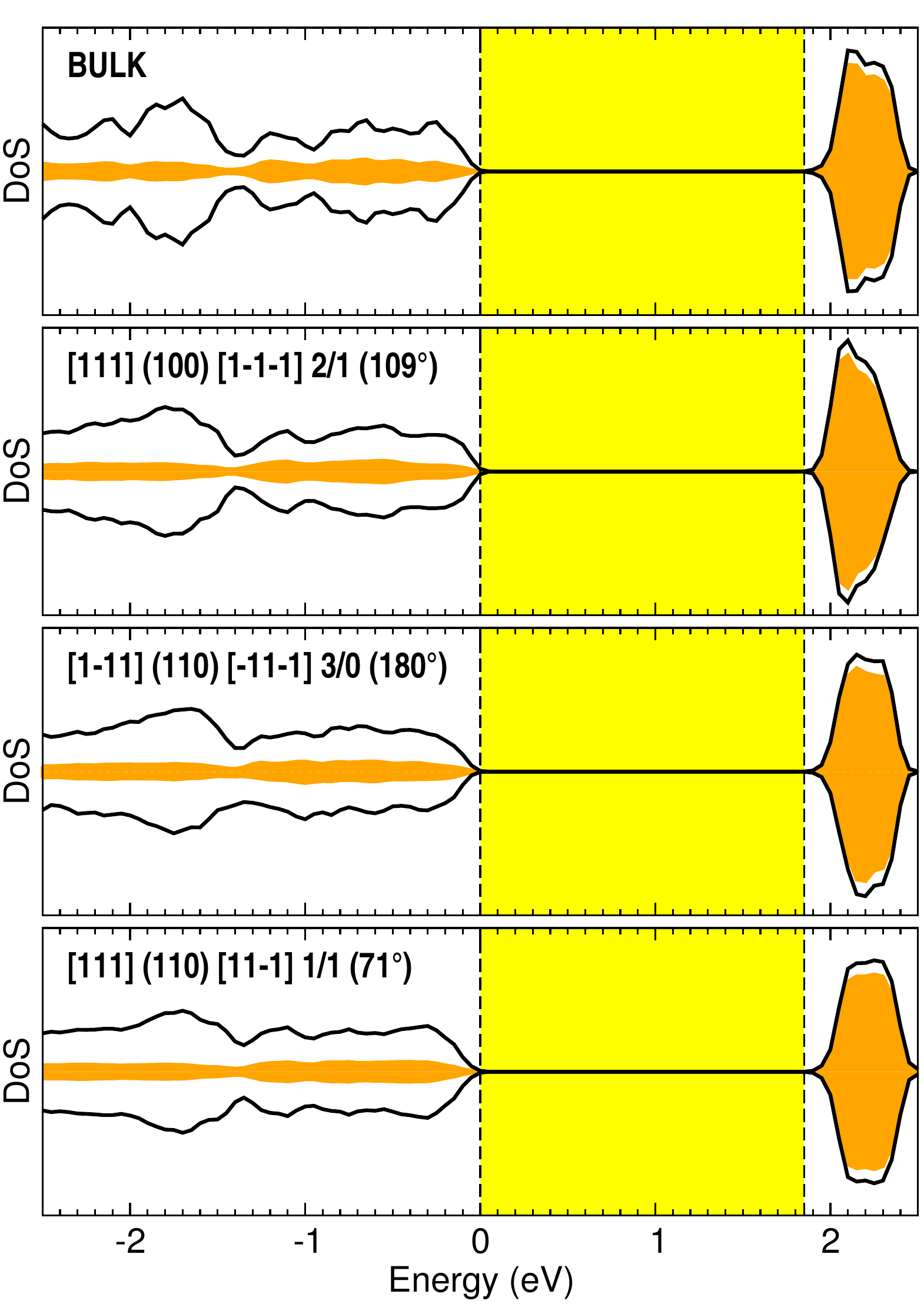}
\caption{(Color online.)  LDoS of bulk BFO and of the three
  representative DW configurations of Table~\ref{tab_lubk_siesta}
  (simulation cells with 80 atoms).  The density of states has been
  projected on the closest 4 Fe, 4 Bi, and 12 O ions to the DW (thick
  lines), and on the $d$ orbitals of those ions (shaded areas).  The
  bandgap is highlighted between two discontinuous vertical lines.  }
\label{fig_ldos}
\end{figure}

Figure~\ref{fig_ldos} shows the LDoS that we obtained from the atoms
that are closest to the DW in each of the configurations of
Tables~\ref{tab_lubk_siesta} and \ref{tab_xc}.
As we can see, if there are
bandgap closings in these structures they are too small to be
captured by this first-principles methodology.  In these conditions,
our interpretation is that conductivity at the DWs is not caused by
any significative band-gap closing resulting from structural changes
at the atomic level near a DW of pure BFO.

\subsection{Connection with experiment}

Seidel {\em et al.}\cite{Seidel2009NM} reported an enhanced
conductivity in FE-DWs of types~2 (109$^\circ$) and~3 (180$^\circ$),
while they found that FE-DWs of type~1 (71$^\circ$) did not conduct in
their experiments. On the other hand, the experiments of Farokhipoor
and Noheda\cite{Farokhipoor2011PRL} revealed an enhanced conductivity
at type~1 FE-DWs as well. The first set of authors argued that the
observed conductivity was consistent with changes in the BFO structure
at the domain wall; their complementary first-principles calculations
showed a small reduction in the bandgap of pure BFO that they
correlated to some degree with the conducting behavior. On the other
hand, Farokhipoor and Noheda concluded that conduction at type~1
FE-DWs was related to the presence of oxygen vacancies. In another
paper, Seidel and coworkers\cite{Seidel2010PRL} also demonstrated that
the wall conductivity in La-doped BFO could be controlled through
chemical doping with oxygen vacancies.

Our results support a mechanism for the conductivity that goes beyond
the (small) effect that DW-specific structural distortions have in the
local electronic properties. As mentioned above, we find negligible
differences between the electronic structure of the DWs and that of
the bulk material. Further, we also investigated the exchange
couplings between iron spins at the DW, in order to check whether a
possible reduction in the magnitude of the magnetic interactions could
have an effect on the spin structure (which might, e.g., be associated with a
reduced N\`{e}el temperature at the DW) and thus on the electronic
structure and conductivity; however, the observed variations never
exceeded a 10~\% of the bulk exchange couplings, suggesting that the
effect in the magnetic (and electronic) structure will be
negligible. Hence, our calculations strongly indicate that the
observed enhanced conductivity is to be linked to {\em extrinsic
  factors} such as oxigen vacancies or other sources of
off-stoichiometry at the DWs. In this sense, we are currently
performing calculations to compute the energy of formation of various
defects at several types of DWs, wanting to determine (i) their effect
in the electronic structure and (ii) their preferencial occurrence in
specific DWs. We believe that this is a promising route to explain the
experimental observations of a relatively large DW conductivity from
first-principles theory.

According to the results in Table~\ref{tab_energies}, the creation of
type~2 and type~3 DWs in bulk, defect-free BFO is favored
energetically over the creation of type~1 DWs. On electrical switching
at high field experiments, Seidel {\em et al.}\cite{Seidel2009NM} were
able to create the three types of domains on 100-nm-thick epitaxial
BFO films grown on a (110) SrTiO$_3$ substrate. On the other hand,
Farokhipoor and Noheda\cite{Farokhipoor2011PRL} mention that their
pulsed-laser deposition films grown on SrTiO$_3$-(001) substrates
showed mostly DWs of type~1. Such a preferential occurrence of
71$^{\circ}$ FE-DWs, which had also been observed in previous
works,\cite{Daumont2010PRB} is in obvious conflict with our preditions
and points at the importance of factors that were not included in our
simulations. Indeed, the experimental results are for thin films grown
in particular orientations and conditions of misfit strain, and in the
presence of a bottom substrate (and electrode) that plays a role as
important as to induce a symmetry breaking in the possible
orientations of the polarization (i.e., from the 8 $\langle 111
\rangle$ polarization variants that can in principle occur, the films
of Ref.~\onlinecite{Farokhipoor2011PRL} only exhibit 4 types of
domains with the polarization always pointing towards the
substrate). As regards the misfit strain, the recent first-principles
investigation of Wei {\sl et al}.\cite{wei-unpublished} suggests that
the DW hierarchy we obtained for the bulk case should be preserved for
films grown on compressive SrTiO$_3$-(001) substrates (with a misfit
strain of about $-$1.5\%). Hence, the discrepancy between our
predictions and the experimental observations seems to indicate that
there are other factors beyond the epitaxial strain -- e.g., the
presence of defects, the effect of interfaces with substrate and/or
electrodes, the specific electric boundary conditions -- that play a
key role in determining which DWs occur spontaneously in BFO films.

Our first-principles results point to a picture of extremely thin
ferroelectric DWs around 1~nm thick, in agreement with previous
calculations for other perovskite oxides. Our ideal
DWs lie on infinite parallel planes, and they do not interact with
each other when they are more than a few nanometers apart. (As
evidenced by the rapid convergence of the computed DW energies as we
increase the separation between DWs in our simulation box.) Instead,
real ferroelectrics show DWs in equilibrium that exhibit a
characteristic roughness,\cite{Paruch2007book} indicating that there
are defects in the lattice that are pinning the
walls.\cite{Catalan2012RMP} In this respect, our simulations can be
taken as the initial step in building a model of ferroelectric domains
that would take into account the role of size effects and defects, a
challenge that is currently out of reach of direct first-principles
calculations.


\section{Summary and conclusions}

We have presented a theoretical study of domain walls (DWs) in
perovskite oxides where two primary order parameters coexist.  We have
taken BiFeO$_3$ as a sample case in which two structural instabilities
of similar strength -- i.e., a ferroelectric (FE) one, and one that
involves concerted rotations of the O$_6$ octahedra
(anti-ferrodistortive or AFD) -- are present in the phase that is
stable at ambient conditions. We have used density-functional theory
to perform atomic scale simulations of the {\em hybrid} domain
boundaries involving both FE and AFD discontinuities, and have thus
revealed several novel aspects that need to be taken into account in
situations like this one. First, and in contrast with what is
implicitly assumed in most BFO studies, we have found that the DW
energetics is essentially determined by the type of discontinuity in
the AFD patterns, and not so much by the change in the FE polarization
accross the wall. Thus, the AFD distortions, which are usually treated
as the {\sl secondary} (i.e., less important) order parameter in BFO,
turn out to play the leading role in determining which DWs are most
stable. Second, we have found that the lowest-energy DWs present
atomic structures that can be directly identified with low-lying
meta-stable BFO phases. Indeed, our results suggest that multi-domain
configurations in BFO are similar to polytypic structures with
(meta-)stable polytypes occurring at the (very thin) domain
boundaries. This finding has an appealing and natural connection with
the novel {\em nano-twinned} structures\cite{Prosandeev2012AFM} that
have been recently proposed to explain several controversial regions
in the phase diagram of BFO and BFO-based solid solutions. Finally, we
have discussed the implications of our results as regards current
experimental work on BFO's DWs, emphasizing the need of considering
various {\em extrinsic factors} to explain some of the most
interesting experimental observations. 


\begin{acknowledgments}
This work has been supported by MINECO-Spain
[Grants
  No. MAT2010-18113, No. MAT2010-10093-E, 
  No. CSD2007-00041, and No. FIS2009-12721-C04-02, 
  the {\em Ram\'on y Cajal} program (OD), and the FPU program 
  (PAP, fellowship AP2006-02958)], and by the 
EC-FP7 project OxIDes (Grant No. CP-FP 228989-2).
OD acknowledges useful
discussions with David Vanderbilt and Karin Rabe during a visit to
Rutgers University made possible by CSIC's i-LINK 2011 program (Grant
No. i-LINK-0438).
Discussions with L. Bellaiche are gratefully acknowledged.
We thankfully acknowledge the computer resources,
technical expertise, and assistance provided by the
Red Espa\~nola de Supercomputaci\'on (RES), and
the Centro de Supercomputaci\'on de Galicia (CESGA).
\end{acknowledgments}




\begin{thebibliography}{99}
\bibitem{fn_reference} Note that, as we move from one domain to
  another, these axes of reference might not exactly coincide, since
  the experimental pseudocubic angle is
  89.20$^\circ$\protect\cite{Kubel1990AC} and not 90$^\circ$. In cases
  where the perovskite structure is strongly distorted, it may be
  convenient to introduce cell-dependent rotation axes, but such a
  choice does not affect the present argument.
\end{thebibliography}
\end{document}